\let\myover=\over   
\documentclass[12pt,a4paper]{article}
\usepackage{amssymb,amsmath,epsfig}
\def\be{\begin{equation}}
\def\ee{\end{equation}}

\def\l{\left(}
\def\r{\right)}

\newcommand{\bg}{\begin{gather}}
\newcommand{\eg}{\end{gather}}

\begin{document}
\let\over=\myover  

\title{\bf Sgoldstino events in top decays at the LHC}

\author{{\em 
   D.~Gorbunov\footnote{{\em e-mail:} gorby@ms2.inr.ac.ru}$^{~,a}$,
   V.~Ilyin\footnote{{\em e-mail:}    ilyin@theory.npi.msu.su}$^{~,b}$,
   B.~Mele\footnote{{\em e-mail:}     Barbara.Mele@roma1.infn.it}$^{~,c}$}\\
   $^a${\small{\em Institute for Nuclear Research of Russian Academy
            of Sciences, Moscow 117312, Russia.}}\\
   $^b${\small{\em Institute of Nuclear Physics, Moscow State University,
            Moscow, 119899, Russia.}}\\
   $^c${\small{\em INFN, Sezione di Roma and Department of Physics, University
            of Rome I, Rome, Italy.}}}

\maketitle

\begin{abstract} 
\small 
  
We study top-quark decays in models with light sgoldstinos. The
off-diagonal entries in the squark mass matrices can lead to FCNC top
two-body decays into a sgoldstino and a $c$ (or $u$) quark. 
We compute the rates
for these decays and discuss the corresponding signatures that could
manifest the presence of sgoldstinos in top decays at the LHC. We expect that a
supersymmetry breaking parameter $\sqrt{F}$  up to a scale of order 10~TeV
could be probed through this process, for a maximal squark mixing with the
third generation. Justified by our preliminary analysis, a thorough study
of the corresponding signal versus background and systematics in the LHC
environment would be most welcome, in order to  accurately assess the
potential of this promising process in determining $\sqrt{F}$.

\end{abstract}

\section{Introduction} 

Among the various supersymmetric extensions of the Standard Model (SM) of
elementary particles, there is a set of models where the goldstino 
superpartners are fairly light, even lighter than the electroweak bosons. 
Such models emerge in the framework of no-scale supergravity~\cite{no-sale}
as well as in gauge mediation (see Ref.~\cite{gmm} and references therein).
These goldstino superpartners --- the scalar $S$ and the pseudoscalar $P$
--- will be referred to as {\it sgoldstinos} in what follows. Their
couplings to the SM fields are governed by ratios of some soft terms from
the superpartner sector and the supersymmetry breaking parameter $\sqrt F$. 
In general, sgoldstino coupling constants  receive contributions from
various terms in the Lagrangian of the underlying theory, but there are
always contributions proportional to soft MSSM terms. In what follows we
will consider only the latter contributions. 

If sgoldstinos are light enough, they may appear in the products of decays
of mesons (e.g., in $J/\psi\to S\gamma$ and $K^+\to S\pi^+$). This issue
was analyzed in Ref.~\cite{sgold-phen}. Current experimental results on the
measurements of meson rare-decay rates place strong bounds on light
sgoldstino couplings. If the supersymmetry breaking soft terms are assumed
to be of the order of the electroweak scale (as motivated by the
supersymmetric solution to the gauge hierarchy problem of the SM), then the
bounds on sgoldstino flavor-conserving coupling constants provide limits on
$\sqrt{F}$ of the order of several hundred GeV. Flavor-violating sgoldstino
interactions might play a more important role in meson physics. The
corresponding coupling constants are proportional to the off-diagonal
entries in the squark soft squared mass matrices. Then, if the off-diagonal
entries are of the order of the current limits coming from the absence of
FCNC, the measurements of mesons decay rates can give bounds on sgoldstino
flavor-violating couplings that are strong enough to probe a scale of
supersymmetry breaking up to $\sqrt{F}\sim 10^7$~GeV~\cite{sgold-phen}. 

Whereas the sgoldstino interactions with quarks of the first and second
generations and with a $b$ quark could be measured in the mesons decays,
the most direct way of determining sgoldstino
flavor-violating  couplings to the top-quark would be through
anomalous top decays (see also~\cite{brig}). 
This approach can be
relevant even for sgoldstino masses  $m_{S(P)}\gtrsim10$~GeV, 
for which the constraints on $\sqrt{F}$ arising 
from the analysis of the meson rare decays become invalid. 

This is indeed the subject of the present letter. We study the anomalous 
top decays into a sgoldstino and an up-type quark $t\to cS(P)$ and $t\to
uS(P)$. Subsequently, depending on the superpartner spectrum and the value
of the $S(P)$ mass, sgoldstinos can decay into SM particles or gravitinos. 
Thus, different final states could be exploited as a signature of a
sgoldstino coupled to the top quark. The sgoldstino masses $m_S$ and $m_P$
that will be relevant for our study are larger than a few GeV's and 
smaller than $m_{top}$.

In order to set the LHC potential, we recall that  $10^7\div 10^8$ top
quarks are expected to be produced with $(10\div 100)$ fb$^{-1}$ of
integrated luminosity at the LHC~\cite{toprev}. Then, on a purely statistical
basis, one will be able to probe top branching ratios (BR) down to about
$10^{-6}\div 10^{-7}$. In the following, we will assume this BR range to be
the LHC {\it statistical threshold} for the observation of a particular top
decay channel. Of course, background problems and systematics will
considerably weaken this potential. Indeed, for some rare top decays that
were analyzed through dedicated Monte Carlo's the corresponding reduction
in the potential has been estimated to be of at least one order of
magnitude~\cite{toprev}.

In this paper, we first compute the top decay BR's  for the channels $t\to
cS, t\to uS, t\to cP, t\to uP$. Then, after reviewing the sgoldstino decay
channels and the corresponding rates, we discuss the possible strategies
for an experimental search for events with a top decaying into sgoldstinos
at the LHC.

\section{Top decay rates into sgoldstinos}

In this section, we work out the width for top decays into sgoldstinos
through FCNC processes. The relevant effective Lagrangian
reads~\cite{top-sgoldstino,top-hadron-colliders}
\begin{equation}
{\cal L}_{eff}=-{\tilde{m}_{U_{13}}^{(LR)2}\over\sqrt{2}F}S\bar{u}t
-{\tilde{m}_{U_{23}}^{(LR)2}\over\sqrt{2}F}S\bar{c}t
-i{\tilde{m}_{U_{13}}^{(LR)2}\over\sqrt{2}F}P\bar{u}\gamma_5t
-i{\tilde{m}_{U_{23}}^{(LR)2}\over\sqrt{2}F}P\bar{c}\gamma_5t\;,
\label{top-sgold-eff}
\end{equation}

\noindent
where $(\tilde{m}_{U_{ij}}^{LR})^2$ are the off-diagonal entries into the
up-squark soft squared mass matrix (for convenience we consider these
parameters to be real~\footnote{In general, there are P-conserving and
P-violating sgoldstino couplings, which are proportional to 
$\tilde{m}_{U_{ij}}^{(LR)2}+(\tilde{m}_{U_{ji}}^{(LR)2})^*$ and 
$\tilde{m}_{U_{ij}}^{(LR)2}-(\tilde{m}_{U_{ji}}^{(LR)2})^*$
respectively (see Ref.~\cite{kaon}). It is enough for our purposes to
consider orthogonal $LR$ mass matrices. The extension to the general
case is straightforward.}). Then, the relevant top decay widths are given by
\begin{equation}
\Gamma(t\to cS(P))=\delta^2_{U_{32}}
{m_t\tilde{m}_U^4\over 32\pi F^2}\l1-{m_{S(P)}^2\over m_t^2}\r^2\;,~
\Gamma(t\to uS(P))=\delta^2_{U_{31}}
{m_t\tilde{m}_U^4\over 32\pi F^2}\l1-{m_{S(P)}^2\over m_t^2}\r^2\;,
\label{rates}
\end{equation}

\noindent
where we ignore the final quark masses and introduce the parameters 
$\delta_{U_{ij}}=(\tilde{m}_{U_{ij}}^{LR})^2/{\tilde{m}_U}^2$, with
$\tilde{m}_U$ being the average mass of up-squarks. These parameters can be
constrained from the absence of FCNC~\cite{masiero} only for the sgoldstino
couplings to the quarks of the first two generations. For example, the
current upper bound on $\delta_{U_{12}}$ at $\tilde{m}_U=M_3=500$~GeV is
about $3\cdot10^{-2}$~\cite{masiero}\footnote{By $M_1$, $M_2$ and $M_3$ we
denote the gaugino masses, the first two for the electroweak groups, and
the third for $SU(3)_c$.}. However, for the couplings to top-squarks there
is no such a constraint yet, and the corresponding value of
$\delta_{U_{3j}}$ (with $j=1,2$) may be as large as 1. 

\begin{figure}[htb]
\begin{center}
{\epsfig{file=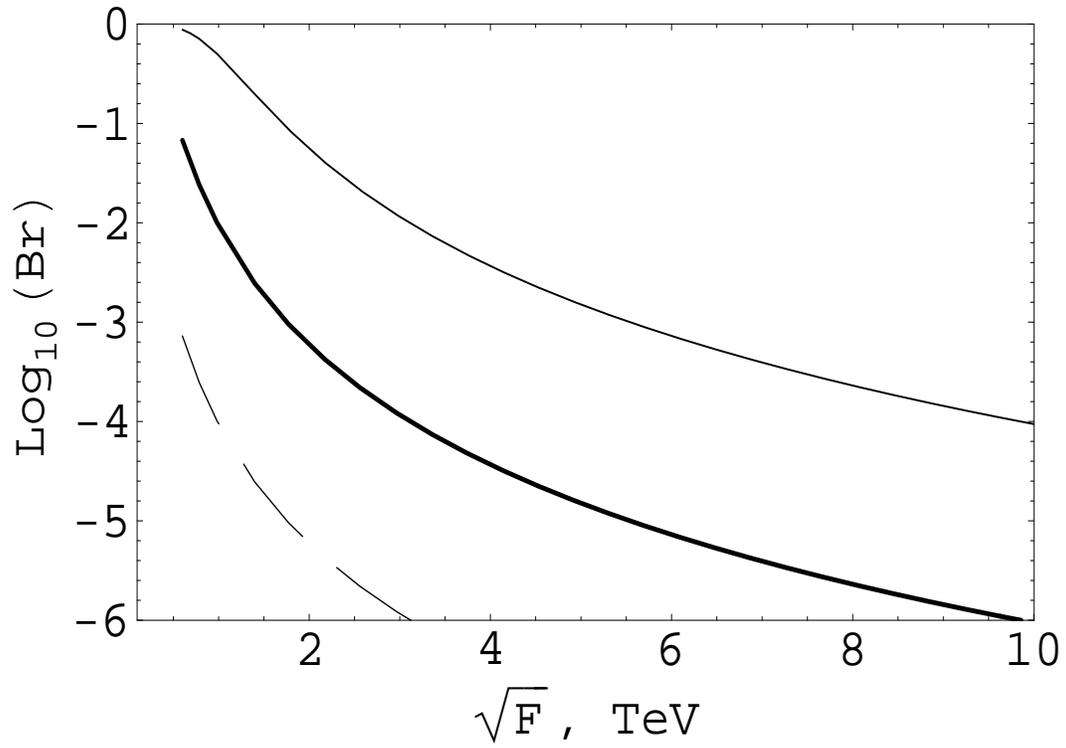,height=10cm,width=15cm}}
\caption{Branching ratio for the top quark decay into a sgoldstino and 
a charm (or up) quark for $\delta_{U_{3j}}=1$ (thin line),
$\delta_{U_{3j}}=10^{-1}$ (thick line) and 
$\delta_{U_{3j}}=10^{-2}$ (dashed line). The average up-squarks mass 
$\tilde{m}_U$ is set to 1~TeV, and 
$m_{S(P)}=50$~GeV.}
\label{rate}
\end{center}
\end{figure}

In Figure~\ref{rate} we present the top BR into  a sgoldstino (either
scalar or pseudoscalar) and a charm-quark (or up-quark) as a function of
$\sqrt{F}$, for an average up-squark mass $\tilde{m}_U=1$~TeV. The three
curves refer to three different values of $\delta_{U_{3j}}$ (i.e.,
$10^{-2}, 10^{-1}, 1)$ . For the sgoldstino mass, we  assume
$m_{S(P)}=50$~GeV, that is a value quite far from where phase-space effects
become important. Only one out of the four 
possible channels $t\to cS, t\to uS, t\to cP,
t\to uP$ is assumed to be open in the computation of the top total
width.

The quartic dependence on $1/\sqrt{F}$ of the width makes the BR for the
top decay into a sgoldstino quite sensitive to the supersymmetry breaking
scale. In the most promising case of maximal mixing (i.e., $\delta_{U_{3j}}
= 1$), one gets BR$(t\to cS) > 10^{-4}$ for $\sqrt{F}$ up to 10 TeV, that
corresponds to more than $10^2$ interesting decays occurring at the LHC for a
wide range of $\sqrt{F}$ (from now on  we refer to each of all the possible
channels $t\to cS, t\to uS, t\to cP, t\to uP$ by simply indicating $t\to
cS$, unless differently specified). On the other hand, decreasing the
mixing $\delta_{U_{3j}}$ by a factor 10 reduces BR$(t\to cS)$ by a factor
$10^2$. For $\delta_{U_{3j}}=10^{-2}$, BR$(t\to cS)$ is more than $10^{-6}$
only up to $\sqrt{F} \sim 3$~TeV, when $\tilde{m}_U=1$~TeV. 

\begin{figure}[htb]
\begin{center}
{\epsfig{file=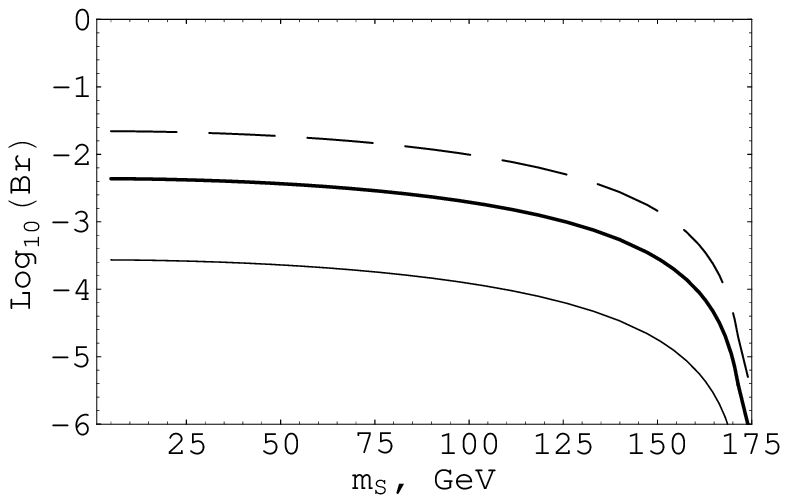,height=10cm,width=15cm}}
\caption{Branching ratio for the top quark decay into a sgoldstino and
a  charm (or up) quark for  $\tilde{m}_U=500$~GeV (thin line),
$\tilde{m}_U=1000$~GeV (thick line) and 
$\tilde{m}_U=1500$~GeV (dashed line). The supersymmetry breaking scale
$\sqrt{F}$ is set to 4~TeV, and $\delta_{U_{3j}}=1$.}
\label{rate2}
\end{center}
\end{figure}

\begin{figure}[htb]
\begin{center}
{\epsfig{file=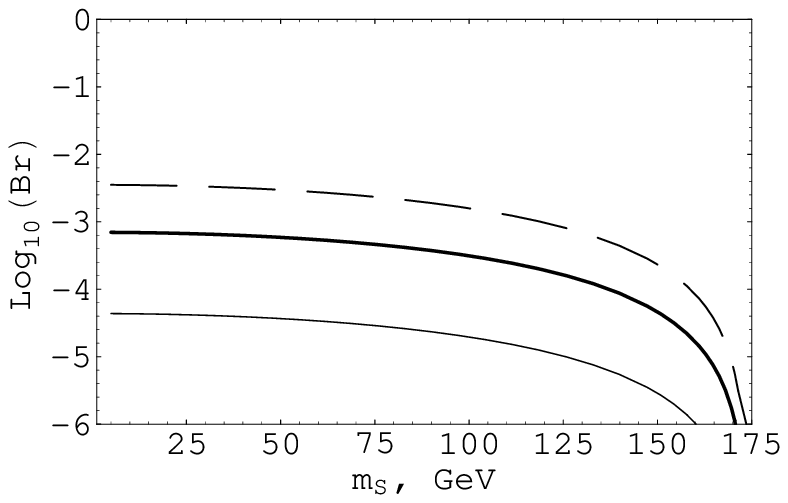,height=10cm,width=15cm}}
\caption{Branching ratio for the top quark decay into a sgoldstino and
a charm (or up) quark for 
 $\tilde{m}_U=500$~GeV (thin line),
$\tilde{m}_U=1000$~GeV (thick line) and 
$\tilde{m}_U=1500$~GeV (dashed line). The supersymmetry breaking scale
$\sqrt{F}$ is set to 2~TeV, and $\delta_{U_{3j}}=0.1$.}
\label{rate3}
\end{center}
\end{figure}

In Figures~\ref{rate2} and \ref{rate3}, we show the phase-space reduction
of BR$(t\to cS)$ for heavier $m_S$ values. We assume, respectively, 
$\sqrt{F}=4$~TeV and $\delta_{U_{3j}}=1$ in Figure~\ref{rate2}, and
$\sqrt{F}=2$~TeV and $\delta_{U_{3j}}=0.1$ in Figure~\ref{rate3}. The three
curves in each figure correspond to different values (i.e., 0.5, 1, 1.5
TeV) for the average up-squark mass. One can see that going to values
$m_S>150$~GeV, a reduction of the BR by more than a factor 10, with respect
to the massless case, makes the problem quite harder in general. On the
other hand, at fixed relative mixing $\delta_{U_{3j}}$, a heavier up-squark
mass $\tilde{m}_U$ enhances BR remarkably. In general, larger values of
$\sqrt{F}$ will be explorable for higher values of $\delta_{U_{3j}}$ and 
$\tilde{m}_U$, and smaller $m_S$. 

\section{Possible signatures at the LHC}

In order to analyze the possible experimental signatures by which a top
decaying into a sgoldstino would manifest at the LHC, one has to consider the
subsequent decay of the sgoldstino inside the experimental apparatus.
Indeed, for the range of parameters that are relevant for this study,
sgoldstinos are expected to decay inside the experimental apparatus, not
far from the collision point~\cite{sgold-modes}. Then, assuming that the
supersymmetric partners (others than the gravitinos $\tilde G$) are too
heavy to be relevant for the sgoldstino decays, the main decay channels are
(for $m_t \gtrsim m_{S(P)}$):  
\begin{equation}
S(P)\to gg, \gamma\gamma, \tilde{G}\tilde{G}, f \bar f, \gamma Z, WW. 
\end{equation}
The corresponding widths have been computed in \cite{sgold-modes}.

For a  sgoldstinos  decaying into a pair of photons, one has
\begin{equation}
\Gamma(S(P)\to\gamma\gamma)={M_{\gamma\gamma}^2m_{S(P)}^3\over 32\pi F^2}\;,
\label{sgold-to-photons}
\end{equation}
\noindent
where $M_{\gamma\gamma}=M_1\cos^2\theta_W+M_2\sin^2\theta_W$, 
and $\theta_W$ is the electroweak mixing angle.

For the two-gluons decay, one similarly finds 
\begin{equation}
\Gamma(S(P)\to gg)={M_3^2m_{S(P)}^3\over 4\pi F^2}\;,
\label{sgoldstino-to-gluons}
\end{equation} 

\noindent
that, for $M_{\gamma\gamma}\sim M_3$,
dominates over the photonic channel due to the color factor enhancement. 

For the $\sqrt{F}$ values we are interested in, gravitinos are very light,
with masses in the range $m_{\tilde{G}}=\sqrt{8\pi/3}\;
F/M_{Pl}\simeq10^{-3}\div 10^{-1}$~eV. Then, the sgoldstino decay rates
into two gravitinos are given by 
\begin{equation}
\Gamma(S(P)\to\tilde{G}\tilde{G})={m_{S(P)}^5\over 32\pi F^2}\;, 
\label{sgold-to-gold}
\end{equation}

\noindent
and become comparable with the rate into two photons for quite heavy
sgoldstinos, such that $m_{S(P)}\sim M_{\gamma\gamma}$.

Sgoldstinos can also decay into fermion pairs, with rates
\begin{equation}
\Gamma(S\to f\bar{f})=N_C{A_f^2m_f^2m_S\over 32\pi
F^2}\l1-{4m_f^2\over m_S^2}\r^{1/2}\;,
~~\Gamma(P\to f\bar{f})=N_C{A_f^2m_f^2m_P\over 32\pi
F^2}\l1-{4m_f^2\over m_P^2}\r^{3/2}\;, 
\label{sgold-to-fermions}
\end{equation} 

\noindent
where $A_f$ is the soft trilinear coupling constant, and $N_C=3$ for quarks
or $N_C=1$ for leptons. One can see that, far from the threshold, the
fermionic BR's are suppressed by a factor $m_f^2/m_S^2$ in general. Hence,
the decay $S(P) \to f \bar f$ can be relevant  for large trilinear
couplings and/or if the sgoldstino mass happens to be not too far from
$m_f$.

Finally, sgoldstinos lighter than the top quark can decay 
into massive vector bosons states. For $m_{S(P)} > M_Z$ and $m_{S(P)} > 2M_W$,
the $Z\gamma$ and  $W^+W^-$ channels open up, respectively. 
The corresponding rates read
\begin{equation}
\Gamma(S(P)\to \gamma Z)={M_{\gamma Z}^2m_{S(P)}^3\over 16\pi
F^2}\l1-{M_Z^2\over m_{S(P)}^2}\r^3\!\!,
~~\Gamma(P\to W^+W^-)={M_2^2m_{P}^3\over 16\pi F^2}
\l1-{4M_W^2\over m_{P}^2}\r^{3/2}\!\!\!\!\!\!,
\label{sgold-to-bosons}
\end{equation}

\begin{eqnarray}
\label{scalar-to-ww}
\Gamma(S\to W^+W^-)={1\over 16\pi F^2m_S}\Biggl( M_2^2\l
m_S^4-4m_S^2m_W^2+6M_W^4
\r-12M_2\mu_aM_W^2\l{m_S^2\over 2}-M_W^2\r\\+2\mu_a^2m_W^4\l
{m_S^4\over4M_W^4}-{m_S^2\over m_W^2}+3\r\Biggr)
\sqrt{1-{4M_W^2\over m_{S}^2}}\;,
\nonumber
\end{eqnarray}  

\noindent
where $M_{\gamma Z}=(M_2-M_1)\cos\theta_W\sin\theta_W$, and the variable
$\mu_a$ that enters $\Gamma(S\to W^+W^-)$ is the higgsino mixing mass
parameter. 

\vspace{0.5cm}

\begin{figure}[htb!]
\begin{center}
{\epsfig{file=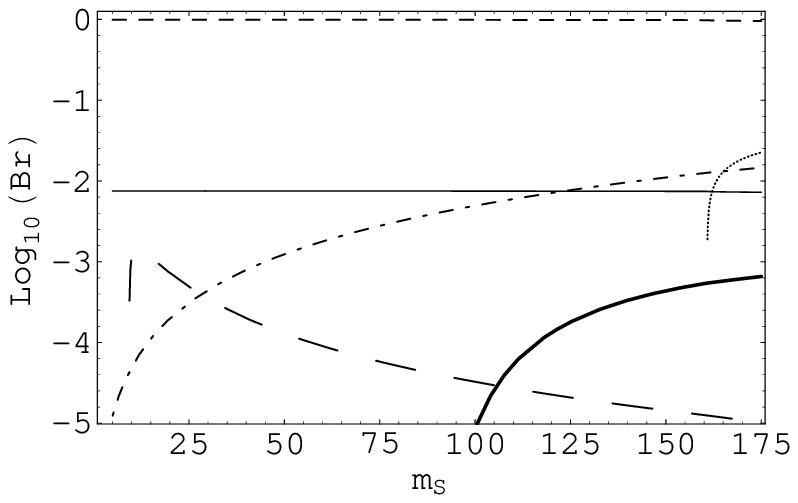,height=5cm,width=7.5cm}}
{\epsfig{file=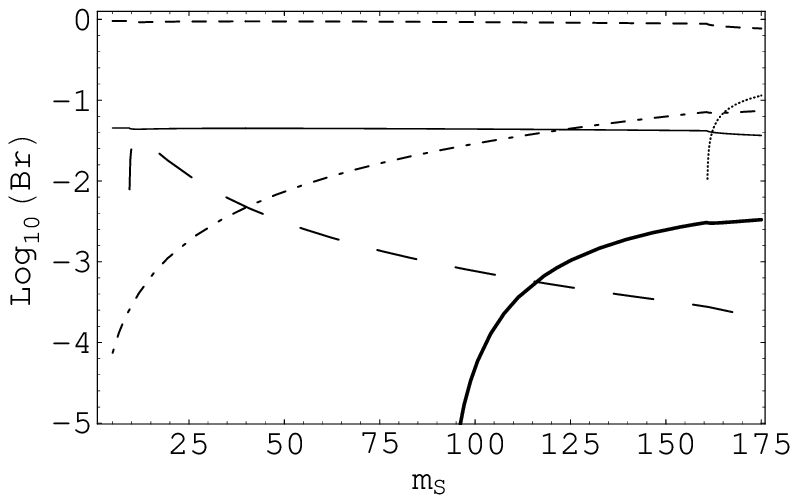,height=5cm,width=7.5cm}}
{\epsfig{file=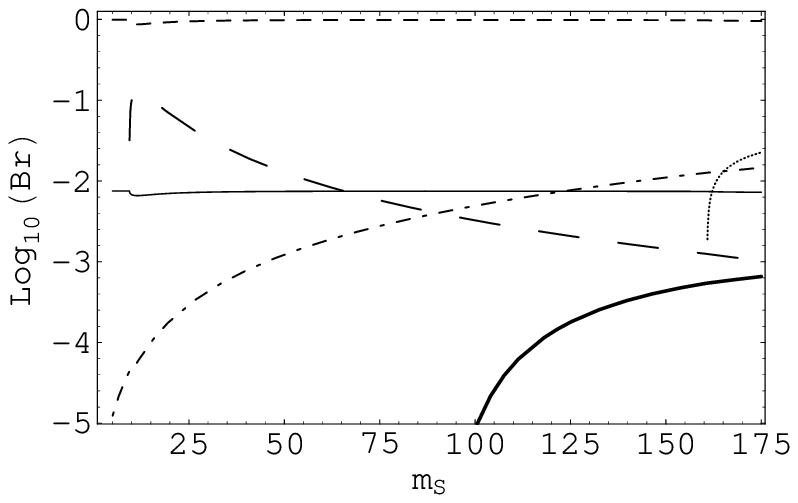,height=5cm,width=7.5cm}}
{\epsfig{file=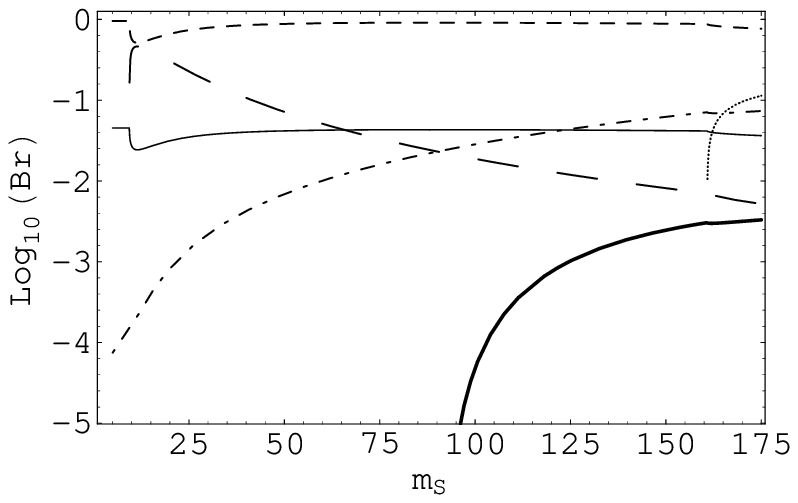,height=5cm,width=7.5cm}}
{\epsfig{file=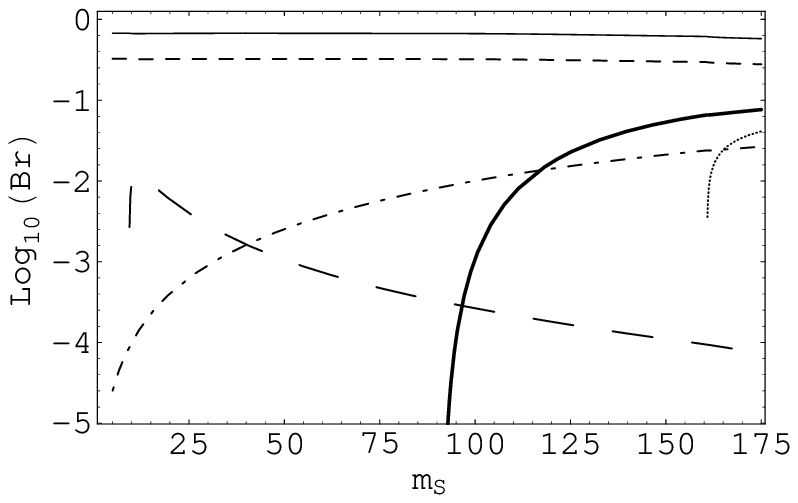,height=5cm,width=7.5cm}}
{\epsfig{file=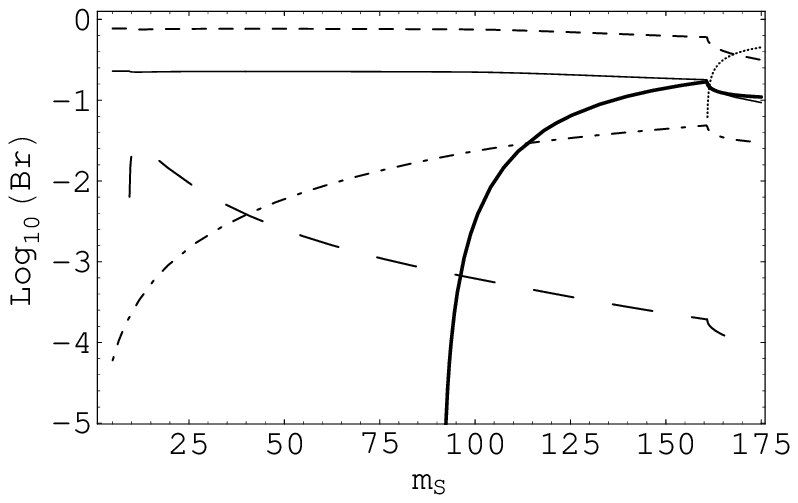,height=5cm,width=7.5cm}}
\caption{Branching ratios for the scalar sgoldstino decays into 
$\gamma\gamma$ (thin line), $gg$ (short-dashed line),
$b\bar{b}$ (long-dashed line), $\tilde{G}\tilde{G}$ (dash-dotted
line), $Z\gamma$ (thick-line) and $W^+W^-$ (dotted line)
in models with $\mu_a=-200$~GeV and:\newline 
a) $M_1=100$~GeV, $M_2=200$~GeV, $M_3=500$~GeV, $|A_b|=100$~GeV;\newline
b) $M_1=100$~GeV, $M_2=200$~GeV, $M_3=200$~GeV, $|A_b|=200$~GeV;\newline 
c) $M_1=100$~GeV, $M_2=200$~GeV, $M_3=500$~GeV, $|A_b|=1000$~GeV;\newline
d) $M_1=100$~GeV, $M_2=200$~GeV, $M_3=200$~GeV, $|A_b|=1000$~GeV;\newline
e) $M_1=1000$~GeV, $M_2=200$~GeV, $M_3=200$~GeV, $|A_b|=200$~GeV;\newline  
f) $M_1=100$~GeV, $M_2=1000$~GeV, $M_3=200$~GeV, $|A_b|=200$~GeV.}   

\label{branchings}
\end{center}
\end{figure}

In Figure~\ref{branchings}, we present the BR's for all the scalar
sgoldstino decay modes for six different sets of the MSSM parameters $M_1$,
$M_2$, $M_3$, $A_b$, and $\mu_a=-200$ GeV. The only channel that is
affected by the parameter $\mu_a$ is $S\to WW$, that, since $m_S>2M_W$, is
anyhow suppressed in the chain $t\to Sc \to WWc$ by the small phase-space
factor in the top decay BR. For this reason, in the Figure~\ref{branchings}
we do not analyze the BR's dependence on $\mu_a$, that was however chosen
negative in order to enhance the corresponding BR($S\to WW$).  

For a typical $M_1,M_2,M_3$ hierarchy, that corresponds to the gauge
unification conditions in the MSSM, one finds that the $S$ decay into
gluons $S\to gg$ always dominates over the other channels. Then, the $S$
decay into photons $S\to \gamma\gamma$ has a BR typically in the range 
$10^{-2}\div10^{-1}$. Decays into $\tilde{G}\tilde{G}$, $\gamma Z$ and $WW$
can be important for relatively large $m_S$ that, however, are disfavored
by the phase-space factor in the top decay BR (cf. Figures~\ref{rate2} and
\ref{rate3})~
\footnote{Note that the three-body decay channels $S\to \gamma Z^*$ 
and $WW^*$ (where  $Z^*$ and $W^*$ are off-shell bosons 
decaying into a {\it real} fermion pair) 
 could be significant in some interval below
the corresponding theresholds for boson pairs, 
similarly to the Higgs boson case.}. 
On the other hand, the sgoldstino decay into $b$ quarks
$S\to b\bar b$ can have BR values in an interesting range when the soft
trilinear coupling $A_f$ is large enough, and for moderate $m_S$. For the
parameter sets of Figures~\ref{branchings}, c) and d), BR($S\to b \bar b$)
reaches values even larger than $10^{-1}$, for $m_S<50$GeV. Particularly
interesting in view of the possibility of detecting a signal in a hadron
collider (see also below) is the case where one drops the typical
$M_1,M_2,M_3$ hierarchy of the MSSM, and allows $M_1$ and/or $M_2$ to be
larger than $M_3$. Representative examples of this case are considered in
Figures~\ref{branchings}, e) and f), where one reaches comparable values of
BR($S\to \gamma\gamma$) and BR($S\to gg$), and even BR($S\to
\gamma\gamma)>$BR($S\to gg$). In the same framework, also the decay $S\to
\gamma Z$ is enhanced up to values of order $10^{-1}$, although part of the
relevant $m_S$ range is penalized by the top  decay phase-space.

The main features of this discussion hold also for the pseudoscalar
sgoldstino $P$. The major difference with respect to the $S$ case lies in
the $P\to WW$ width (that is independent of $\mu_a$). Again, this channel is
relevant only for $m_P$ values penalized by the top decay phase space. 

\vspace{0.6 cm}

We now discuss the relevance of the different $S$ decay channels in view of
the possible detection of a $t\to Sc$ signal at the LHC. 
A complete analysis will
require a thorough study via Monte Carlo's that can treat each particular
signal versus the complicated backgrounds and systematics of a hadronic
machine.

Here, we will work out a few {\it qualitative} 
conclusions, based on the results of previous dedicated studies that set
the threshold for the observation of top rare decays and Higgs decays at
the LHC~\cite{toprev,atlas,j2g}. These results sum up in the following
assumptions for our analysis:

\vspace{0.3cm}

{\it i)} we require the production of 10 events for a given signature (or
particular decay of $S$ following $t\to Sc$) to be observable. On a purely
statistical basis, this corresponds to a {\it total} top BR in that channel,
given by the product BR($t\to Sc)\cdot$BR($S\to...)$, in the range
$10^{-6}\div 10^{-7}$, for $10^7\div 10^8$ top quarks produced at the LHC. 

\vspace{0.3cm} 

{\it ii)} we assume that a generic {\it electroweak-like signature} 
(like the ones involving photons, $Z$'s, $W$'s and/or missing momentum) can
be detected at the LHC, and measured with an efficiency of the order of $10\%$
due to the background subtraction and systematics. The same efficiency is
assumed for the detection of a resonant $b\bar b$ system.

\vspace{0.3cm} 

{\it iii)} we assume that a completely hadronic (non $b$-like) signature of
a top rare decay will not be manageable at the LHC because of the QCD
backgrounds. This excludes from our study the $S(P)\to gg$ channel.

\vspace{0.3cm} 
In this framework, in order to be able to detect a sgoldstino decaying into
the channel $S\to X$ at the LHC, one needs 
\begin{equation}
      \mathrm {BR}(t\to Sc)\cdot \mathrm {BR}(S\to X) 
      \gtrsim \mathrm {BR}_{th} = 10^{-5}
\label{sg-th}
\end{equation}
and ``$X$" giving rise to either an 
{\it electroweak-like signature} or a $b\bar b$
system.

\vspace{0.5cm} 

Let's go through some details on the detection of the possibly relevant $S$
decay modes:

\begin{itemize}

\item{$S(P)\to \gamma\gamma$} 

This sgoldstino decay mode gives rise to the total final state $t\bar t\to
(W+b$-$jet)+(\gamma\gamma+jet)$ in top pair production at the LHC. One should
tag one of the top in the usual $(W+b$-$jet)$ way. Then one has a hard
hadronic jet plus a resonant two-photon system, with a total invariant mass
of the $(2\gamma+jet)$ system of about $m_t$. A proper cut on the
$\gamma\gamma$ invariant mass can help in suppressing the background
through similar strategies to the ones worked out for the $H\to
\gamma\gamma$ detection at the LHC~\cite{atlas}. Assuming the threshold rate of
Eq.(\ref{sg-th}) and a typical BR($S(P)\to\gamma\gamma) \sim
10^{-2}\div10^{-1}$, one can explore through this very distinctive
signature a value for BR($t\to Sc$) down to 
$10^{-3}\div10^{-4}$. In particular, for a maximal squark mixing and
$\tilde{m}_U=1$~TeV, one can probe $\sqrt{F}$ up to about 10 TeV (cf.
Figure~\ref{rate}). For a squark mixing $10^{-1}$ one can cover up to
$\sqrt{F}\sim 3$~TeV. The dedicated assumptions of a broken $M_1,M_2,M_3$
hierarchy in Figures~\ref{branchings}, e) and f), would translate into even
more promising ranges, with BR($t\to Sc$) explorable down to $10^{-5}$, and,
correspondingly, into larger values of $\sqrt{F}$.

\item{$S(P)\to b \bar b$} 

In this case, the total
final signature is $t\bar t\to (W+b$-$jet)+(2 b$-$jets +jet)$
 with a resonant $2 b$-$jets$ system. 
 For large soft
trilinear couplings $A_f$ and  moderate $m_S$ (i.e., less then about 50 GeV), 
one could explore BR($t\to Sc$) ranges down to $10^{-3}\div10^{-4}$,
assuming the threshold in Eq.(\ref{sg-th}).
Hence, this channel could have a comparable potential to the 
$S(P)\to \gamma\gamma$ one, but for a restricted  $m_S$ range.

\item{$S(P)\to \tilde G\tilde G$} 

Gravitinos give rise to missing transverse energy in the event. The total
final signature is then $(W+b$-$jet+jet+E^{mis}_t)$. Assuming that the
threshold in Eq.(\ref{sg-th}) is also effective in this case, this channel
can be useful for intermediate values of $m_S$ in the range
$100\div150$~GeV. Indeed, the relevant $S$ width rises as the fifth power
of $m_S$ and is typically of the order $10^{-2}$ or more in this range.
Larger values of $m_S$ are penalized by the phase-space factor in the top
decay BR. 
Hence, through this channel, one can hope to investigate  
top BR's into a sgoldstino down to
$10^{-3}$ in the $m_S$ range $100\div150$~GeV.

\item{$S(P)\to \gamma Z$}

This channel has a nice signature ($W+b$-$jet+Z \gamma + jet$), with a
resonant $\gamma Z$ system. On the other hand, BR($S\to \gamma Z$) is
hardly larger than $10^{-3}$, unless one breaks the gauge-unification
hierarchy, assuming $M_1$ and/or $M_2$ larger than $M_3$, [cf.
Figures~\ref{branchings}, e) and f)]. In the latter case, one can have
BR($S\to \gamma Z)\sim 10^{-1}$ at $m_S >120$~GeV, with a corresponding
reach of $10^{-4}$ for the top BR  in the $m_S$ range $120\div150$~GeV.

\item{$S(P)\to WW$}

This channel can be interesting from the point of view of the BR($S(P)\to
WW$) value that can easily reach the $10^{-1}$ level. On the other end, the
complete signature ($W+b$-$jet+W+W+jet$) is hardly viable, since it is
relevant for a $m_S$ range ($m_S >160$GeV) where the phase-space
suppression of the top width is fully effective.

\end{itemize}

Note that one could also exploit the single-top production sample
of top quarks, with a different final signature. This would enlarge
the present statistics by about 15\%-20\% \cite{toprev}. 

\section{Conclusions}

We considered the possibility to detect at the LHC anomalous top decays into
light sgoldstinos, proceeding through off-diagonal flavor-violating entries
in the squark-mass matrices. We studied both the decay rates and the
relevant experimental signatures. The most promising signatures are the
ones associated with the decay chains $t\to qS \to q \gamma \gamma$ and
$t\to qS \to q b \bar b$, where $q$ is either a $c$ or a $u$ quark. The
first can be effective for the complete $m_S$ range (from a few GeV up to
about 150 GeV), while the latter can be useful for rather light
sgoldstinos. Assuming a total 10\% efficiency in the detection of these
signatures versus backgrounds and systematics, BR($t\to qS$) values can be
studied down to $10^{-4}$, on the basis of the $10^7\div 10^8$ top quarks
produced with a $(10\div 100)$ fb$^{-1}$ integrated luminosity at
the LHC~\cite{toprev}.

For a maximal flavor-violating LR-mixing of the stop with the other up-type
squark, one could than explore the scale of supersymmetry breaking
$\sqrt{F}$ up to about 10 TeV, that is, e.g., much further than what is
reachable in searches for light gravitinos at the LHC~\cite{gra-lhc}. A
dedicated and thorough analysis via Monte Carlo techniques will be required
in order to accurately assess the potential of this promising process.

\vspace{2cm}
{\large \bf Acknowledgments} 
\vspace{0.5cm}

The work of D.G. was supported in part under RFBR  grant 99-01-18410, CRDF
grant (award RP1-2103), by the the Council for
Presidential Grants and State Support of Leading Scientific Schools, grant
00-15-96626, by the program SCOPES of the Swiss National Science
Foundation, project 7SUPJ062239. 
The work of V.I. and B.M. was supported in part
by the CERN-INTAS 99-0377 grant.

\def\ijmp#1#2#3{{\it Int. Jour. Mod. Phys. }{\bf #1~} #3 (19#2)}
\def\pl#1#2#3{{\it Phys. Lett. }{\bf B#1~} (19#2) #3}
\def\zp#1#2#3{{\it Z. Phys. }{\bf C#1~} #3 (19#2)}
\def\prl#1#2#3{{\it Phys. Rev. Lett. }{\bf #1~} #3 (19#2)}
\def\rmp#1#2#3{{\it Rev. Mod. Phys. }{\bf #1~} #3 (19#2)}
\def\prep#1#2#3{{\it Phys. Rep. }{\bf #1~} (19#2) #3}
\def\pr#1#2#3{{\it Phys. Rev. }{\bf D#1~} (19#2) #3}
\def\np#1#2#3{{\it Nucl. Phys. }{\bf B#1~} (19#2) #3}
\def\mpl#1#2#3{{\it Mod. Phys. Lett. }{\bf #1~} #3 (19#2)}
\def\arnps#1#2#3{{\it Annu. Rev. Nucl. Part. Sci. }{\bf #1~} #3 (19#2)}
\def\sjnp#1#2#3{{\it Sov. J. Nucl. Phys. }{\bf #1~} #3 (19#2)}
\def\jetp#1#2#3{{\it JETP Lett. }{\bf #1~} #3 (19#2)}
\def\app#1#2#3{{\it Acta Phys. Polon. }{\bf #1~} #3 (19#2)}
\def\rnc#1#2#3{{\it Riv. Nuovo Cim. }{\bf #1~} #3 (19#2)}
\def\ap#1#2#3{{\it Ann. Phys. }{\bf #1~} #3 (19#2)}
\def\ptp#1#2#3{{\it Prog. Theor. Phys. }{\bf #1~} #3 (19#2)}
\def\spu#1#2#3{{\it Sov. Phys. Usp.}{\bf #1~} #3 (19#2)}
\def\apj#1#2#3{{\it Ap. J.}{\bf #1~} #3 (19#2)}
\def\epj#1#2#3{{\it Eur.\ Phys.\ J. }{\bf C#1~} #3 (19#2)}
\def\pu#1#2#3{{\it Phys.-Usp. }{\bf #1~} (19#2) #3}

\newpage
{\Large \bf Figures}


\begin{figure}[htb!]
\begin{picture}(0,0)(-28,-86)
\put(0.00,277.00){\makebox(0,0)[cb]{e)}}
\put(0.00,420.00){\makebox(0,0)[cb]{c)}}
\put(0.00,564.00){\makebox(0,0)[cb]{a)}} 
\put(220.00,420.00){\makebox(0,0)[cb]{d)}}
\put(220.00,564.00){\makebox(0,0)[cb]{b)}} 
\put(220.00,277.00){\makebox(0,0)[cb]{f)}}
\end{picture}
\end{figure}


\end{document}